# A Vision of C-V2X: Technologies, Field Testing and Challenges with Chinese Development

Shenzhi Chen, *Fellow, IEEE*, Jinling Hu, Yan Shi, Li Zhao, and Wen Li

*Abstract*—C-V2X (Cellular Vehicle-to-Everything) is the important enabling technology for autonomous driving and intelligent transportation systems. It evolves from LTE (Long Term Evolution)-V2X to NR (New Radio)-V2X, which will coexist and be complementary with each other to provide low latency, high reliability, and high throughput communications for various C-V2X applications. In this article, a vision of C-V2X is presented. The requirements of the basic road safety and advanced applications, the architecture, the key technologies, and the standards of C-V2X are introduced, highlighting the technical evolution path from LTE-V2X to NR-V2X. Especially, based on the continual and active promotion of C-V2X research, field testing and development in China, the related works and progresses are also presented. Lastly, the trends of C-V2X applications with technical challenges are envisioned.

## I. INTRODUCTION

CELLULAR Vehicle-to-Everything (C-V2X) is the V2X communication technology developed based on the cellular systems. It leverages and enhances the current cellular network features and elements to enable low-latency and high-reliability communications among various nodes in vehicular networks, including vehicle-to-vehicle (V2V), vehicle-to-pedestrian (V2P), vehicle-to-infrastructure (V2I), and vehicle-to-network (V2N) communications [1]. C-V2X has attracted lots of interests of academic and industrial experts from information technology, automotive engineering and transportation engineering.

With the evolution of cellular systems from 4G LTE to 5G, C-V2X evolves from LTE-V2X to NR-V2X [1]. In May 2013, based on TD-LTE (Time Division Long Term Evolution), CATT (China Academy of Telecommunication Technology) /Datang firstly proposed LTE-V technology in the world, including two modes: LTE-V-Direct (Decentralized mode) and LTE-V-Cell (Centralized mode) [2]. LTE-V-Direct can provide direct V2V/V2I communications and support the basic road safety applications with low-latency, high-reliability, and enough long range. Utilizing eNB (evolved Node B) as the central control node and the data forwarding node, LTE-V-Cell can improve the success access rate and the system capacity with the centralized scheduling scheme, congestion control and interference coordination with the reporting information from the vehicles. Combining the two modes of LTE-V-Direct and LTE-V-Cell, LTE-V can achieve significant operational benefit and spectrum efficiency. And later, LTE-V completed the standardization as LTE-V2X in 3GPP (The 3rd Generation Partnership Project) Rel-14, which aims to provide information exchanging capability for the basic road safety services. NR-V2X is proposed with 5G NR (New Radio), which will coexist with LTE-V2X to support advanced applications such as vehicle platooning, autonomous driving, remote driving, and extended sensors etc.[3][4].

C-V2X is actually the important enabling technology towards autonomous driving and intelligent transportation system. Indeed, China has been actively pushing the progress of C-V2X. The Chinese companies, such as CATT/Datang and Huawei, have been promoting the standardization, technology research and development, as well as testing and industrial practice works of C-V2X in recent years. The cooperation with the automobile and transportation industries have been promoted to support the applications using C-V2X technology.

Therefore, in this article, a vision of C-V2X will be presented. The rest of this article is organized as follows. The technologies, standards and global development of C-V2X are introduced in Section II. Section III focuses on the active promotion works for C-V2X in China. In Section IV, the technical trends with analysis on challenges are presented. Lastly, the article is concluded in Section V.

## II. C-V2X: FROM LTE-V2X TO NR-V2X

The initial version of C-V2X is LTE-V2X, which is primarily based on LTE and can be enough for basic road safety services. Through LTE-V2X, the status information (such as position, heading and speed etc.) can be exchanged among neighboring vehicles, pedestrians, and infrastructures. With the evolution of standards and technologies, LTE-V2X is complemented and This work was supported in part by the National Science Foundation Project in China under grant 61731017, 61731004 and 61931005. *(Corresponding Author: Shanzhi Chen.)*

S. Chen is with the State Key Laboratory of Wireless Mobile Communications, China Academy of Telecommunication Technology, Beijing 100191, China, and also with the State Key Laboratory of Networking and Switching Technology, Beijing University of Posts and Telecommunications, Beijing 100876, China (e-mail: chensz@cict.com).

J. Hu, L. Zhao and W. Li are with the State Key Laboratory of Wireless Mobile Communications, China Academy of Telecommunication Technology, Beijing 100191, China (e-mail: hujinling@catt.cn; zhaoli@catt.cn; liwen@datangmobile.cn).

Y. Shi is with the State Key Laboratory of Networking and Switching Technology, Beijing University of Posts and Telecommunications, Beijing 100876, China (e-mail: shiyan@bupt.edu.cn).

not to be replaced by NR-V2X to support the current and future V2X applications with enhanced radio layer technical features and network architecture. Using the active communications to enable the enhanced perception horizon over varied use cases, C-V2X can comprise the comprehensive system to support the various V2X applications and a multitude of V2X scenarios.

*A. Technologies*

With the development of the communication technologies, C-V2X can pave the clear evolution path from LTE-V2X to NR-V2X. The following key technologies have been researched and developed during the evolution of C-V2X standards and technologies.

*1) The requirements of the basic road safety and the advanced V2X services*

The basic road safety services are supported by 3GPP Rel-14/Rel-15 sufficiently to broadcast the warning messages to the nearby entities [3]. In order to reflect the more stringent requirements and advanced functional aspects of technologies, the 25 use cases have been selected for the advanced V2X services and categorized into four groups with vehicles platooning, extended sensors, advanced driving, and remote driving [4]. The performance requirements for the applications supported by 3GPP Rel-14/Rel-15 and the four groups of advanced V2X services are summarized in Table I.

*2) The centralized/ distributed architecture and communications*

To accelerate the LTE-V2X development, LTE-D2D (Device-to-Device) is reused for LTE-V2X with the enhancements to adapt to the requirements of V2X applications. LTE-V2X reuses the centralized scheduling mode (Mode 3) and decentralized scheduling mode (Mode 4) of LTE-D2D to support the direct communications in the V2X sidelink [5]. In order to operate the network control functionality of authorization and provisioning, the new V2X Control Functions are added into the legacy LTE core network [6].

In IEEE 802.11p, CSMA/CA (Carrier Sense Multiple Access/Collision Avoidance) scheme is used for the resource allocation with the contention overhead [7], while in LTE-V2X, the SPS (Semi-Persistent Scheduling) scheme assisted with the sensing results is selected as the resource allocation mechanism. The SPS mechanism of LTE-V2X adapts to the periodic attributes of the basic safety services, and the counter-based reselection scheme is utilized to avoid the continuous resource collision caused by the half-duplex adverse impact. With the sensing results, the occupation of the resources can be estimated in the resource selection window. According to the priorities of the services, the different sensing thresholds are configured to reflect the employment of the resources. With the higher priority, the packets will have more opportunities to transmit in the sidelink [5]. For Mode 3, SPS scheduling and dynamic scheduling are supported with the reporting information from the OBUs (On Board Units) and RSUs (Road Side Units). Up to eight SPS processes can be configured and activated simultaneously to decrease the scheduling overhead of eNB. For the event-triggered messages, the dynamic scheduling can be used to adapt to the variable messages [8][9].

*3) Key Technologies of C-V2X*
• The physical layer structure design

Considering the high speed (140 km/h) and high center frequency (5.9 GHz), the DMRS (Demodulation Reference Signal) density is increased from two to four columns in time domain for LTE-V2X in the high Doppler case to satisfy the channel estimation and synchronization tracing.

The higher reliability at the same range can be achieved by LTE-V2X with the coding gain from turbo codes, HARQ (Hybrid Automatic Repeat Request), and SC-FDM (Single Carrier Frequency Division Multiplexing Access). With the synchronous scheme and FDM (Frequency Division Multiplexing) in resource allocation scheme of LTE-V2X, the spectral efficiency and the system capacity can be improved.

• Resources allocation and management of C-V2X

When the receiving UE wants to receive data from the distant transmitters and the nearby UEs using the overlapped resources to transmit data, the significant Near-Far adverse impact may happen at the receiving UE to fail decoding the packets from the distant transmitters. The geographical information of zones can be utilized for eNB in centralized mode and for UEs in distributed mode. The resources of different zones can be allocated by eNB and UEs to avoid resource collision and provide the potential spatial division multiplexing gain [5][9].

TABLE I
THE PERFORMANCE REQUIREMENTS FOR THE BASIC ROAD SAFETY AND ADVANCED V2X SERVICES

| Use case group | Transmission mode | Latency (ms) | Reliability | Maximum Data Rate (Mbps) | Communication Range (m) |
|---|---|---|---|---|---|
| Basic road safety services supported by 3GPP Rel-14/Rel-15 | Broadcast | 10-100 | 90% | 31.7 | 100-300 |
| Vehicles Platooning | Broadcast, groupcast and unicast | 10-25 | 90% - [99.99%] | [65] | less than 100m; [5-10] sec * max relative speed |
| Advanced driving | Broadcast | [3-100] | [99.99%]-[99.999%] | [50] | [5-10] sec * max relative speed |
| Extended sensor | Broadcast | 3-100 | [90%-99.999%] | 1000 | [50-1000] |
| Remote driving | Unicast | [5-20] | [99.999%] | UL: 25 DL: 1 | Same as cellular uplink and downlink |

- NR-V2X potential enhancements

LTE-V2X can cover as much as eV2X (enhanced V2X) services and satisfy the basic requirements of some platooning and limited automated driving applications. NR-V2X can be complementary to LTE-V2X and supports more advanced services especially when the requirements cannot be supported by LTE-V2X. The NR-V2X system is expected to have flexible design to support the stringent requirements of V2X services of extreme low latency, high reliability, high system capacity, better coverage, and easy extension to the future deployment and development [11], [12].

Based on the 5G new radio interface design, the physical layer structure of sidelink signals, channels, bandwidth parts, and resource pools are proposed to support more transmission types (unicast, groupcast) with available feedback besides broadcast. Based on the 5G NR design, the enhanced resource allocation scheme to support the advanced V2X services and synchronization scheme are under development. With the 5G novel core network architecture and technical features (MEC (Mobile Edge Computing), NFV (Network Function Virtualization)/SDN (Software Defined Network), and Network slicing), QoS (Quality of Service) management can be realized to satisfying stringent and varied service requirements. The co-existence issues of LTE-V2X and NR-V2X have been considered to provide the inter-operability during the network evolution. Thus, the flexible selection mechanism of LTE-V2X and NR-V2X can be based on the requirements of applications and QoS.

Based on the introduction of the LTE-V2X and NR-V2X technologies, the key differences among IEEE 802.11p, LTE-V2X and NR-V2X are presented in Table II. Therefore, harmonization can pave the evolution path from LTE-V2X to NR-V2X and maintain the advantages over IEEE 802.11p.

*B. Standards*

The V2X standards can be classified into two parts of radio and application. The radio layer standards have been focused on in 3GPP, and the higher layer standards are developed by countries and regions to reflect the realistic and differentiated requirements of V2X applications.

Based on the research progress of LTE-V2X related technologies, the development of LTE-V2X standards has been completed, and the research and development of NR-V2X is in progress by the international and national standardization organizations. The V2X technologies developed in 3GPP with technical specifications will be endorsed by the International Telecommunication Union (ITU) or ISO (International Organization for Standardization) as V2X international standards. These 3GPP technical specifications are then transposed into standards by the regional Standards Setting Organizations (SSOs), such as CCSA (China Communications Standards Association) in China. The regional Standards Developing Organization have conducted V2X projects to promote the related technologies and standards.

TABLE II
THE KEY DIFFERENCES AMONG IEEE 802.11P, LTE-V2X AND NR-V2X [2][5][7][8][9]

|  | IEEE 802.11p | LTE-V2X (3GPP Rel-14/Rel-15) | NR-V2X (3GPP Rel-16/Rel-17) |
|---|---|---|---|
| Standardization Completion | Completed in Mar. 2012 | Rel-14: Mar. 2017 Rel-15: Jun. 2018 | Rel-16: Dec. 2019 Rel-17: Jun. 2021 |
| Evolution path | Forward compatible to IEEE 802.11bd | Forward compatible to NR-V2X | Backward compatible to LTE-V2X |
| Network coverage support | Limited, through AP to connect with network | Yes | Yes |
| Out of network operation | Yes | Yes | Yes |
| Latency | No deterministic delay | Rel-14: 20ms Rel-15: 10ms | 3ms or lower |
| Reliability | No guaranteed reliability | Rel-14: > 90% Rel-15: > 95% | 99.999% |
| Data rate | 6 Mbps | 30 Mbps | Not determined |
| Synchronization | No | Yes | Yes |
| Waveform | OFDM | SC-FDM | CP-OFDM |
| Channel coding | Convolutional | Turbo | LDPC |
| Resource mutiplexing | TDM only | TDM and FDM | TDM and FDM |
| HARQ | No | Yes, fixed two transmissions | Yes, flexible transmissions |
| Resource allocation | CSMA/CA | Sensing and SPS | Not determined |
| Multi-antenna | UE implementation | Transmission diversity, 2Tx/2Rx | Not determined |
| Modulation | 64QAM | 64QAM | 256QAM |

AP: Access point
OFDM: Orthogonal Frequency Division Multiplexing
CP-OFDM: Cyclic-Prefix Orthogonal Frequency Division Multiplexing
LDPC: Low-density Parity-check Code
TDM: Time Division Multiplexing
64QAM: 64 Quadrature Amplitude Modulation

*1) 3GPP*

With the evolution of 4G/5G cellular network communication technologies, C-V2X standardization is driven by 3GPP and divided into three phases. The timeline of 3GPP C-V2X standardization is shown in Fig. 1.

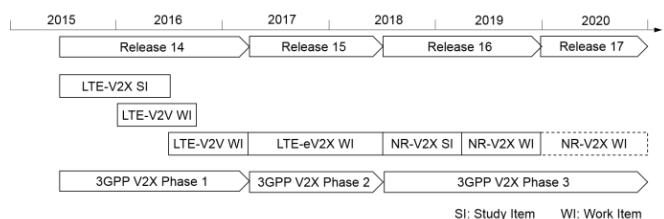

Fig. 1. The timeline of 3GPP C-V2X standardization

Completed in March 2017, the phase 1 of Rel-14 provides the initial standards supporting V2V services and the additional V2X services leveraging the cellular infrastructure. The basic safety services can be delivered in Rel-14 with the exchanging of the status information from neighboring nodes through sidelink communications. In order to support the enhanced V2X services in holistic and complementary manner to Rel-14 LTE-V2X, the phase 2 of LTE-V2X as the Rel-15 is completed in June 2018 to build a strong and stable ecosystem around LTE-V2X. The enhanced technical features in sidelink are introduced in Rel-15, such as carrier aggregation (up to eight component carriers), higher order modulation (64QAM), latency reduction (minimum as 10ms). The feasibility study on both transmission diversity and short TTI is also included. Rel-14 established the foundation for basic safety services. Rel-15 can provide better link budget leading to longer range, lower latency and higher reliability.

Evolution to NR-V2X, the backward compatibility with Rel-14 and Rel-15 can be achieved for NR-V2X. The new NR-V2X system should be designed to support the advanced V2X services with the stringent requirements of lower latency, higher reliability, and higher throughput. Meanwhile, the higher system capacity, larger coverage, extendibility to support future V2X services should be satisfied in the NR-V2X system design. The standardization of NR-V2X has been launched in June 2018, and Rel-16 can be completed as planned and Rel-17 are expected to be completed in the December 2019 and June 2021 [11], [12].

The Chinese enterprises such as CATT/Datang and Huawei have led and actively taken part into the research and development of the standardization of LTE-V2X and NR-V2X.

*2) ITU*

In 2019, WRC (World radiocommunication conferences)-19 1.12 will discuss worldwide or regional harmonization of frequencies for ITS applications. The SG (Study Group)-17 cooperative security will be researched and developed in ITU-T.

*3) ISO*

ISO is a worldwide federation of national standard bodies. In April 2017, the standardization of LTE-V2X is approved in ISO TC204. LTE-V2X is selected as the ISO candidate technologies to support ITS services. The related standard works are still in progress as ISO/CD (Committee Draft) 17515-3 [13].

*4) ETSI (European Telecommunications Standards Institute)*

ETSI TC ITS (ETSI Technical Committee Intelligent Transport Systems) is in charge of developing standards related to the overall V2X communication architecture, management, and security. DSRC (Dedicated Short Range Communications) related protocols have been developed as the ITS-G5 communication standards with the physical and medium access layer technologies of IEEE 802.11p in ETSI. In order to provide the C-V2X communications, the C-V2X standardization process is speeded up in 2018 and the related C-V2X standards work has been completed. The access layer, network and transport layer, and the application layer protocols of C-V2X in ETSI have been defined to provide the availability of the C-V2X protocol stack.

*5) SAE (Society of Automotive Engineers)*

To accelerate the C-V2X standardization and industrial progress, the C-V2X Technical Committee of SAE was formed in June 2017. To adapt SAE J2945.1 to LTE-V2X, SAE J3161 addresses the system requirements and desired interoperability, and provides the standards profiles, functional parameters, and performance requirements [14]. The related standard work has been completed in 2018.

*C. Global Development*

5GAA (5G Automotive Association), as a global and cross-industry organization, was created on Sep. 2016 and has presented the clear position of supporting C-V2X as the feasible solutions for the mobility and transportation services [15]. More than 110 companies from the automotive, telecommunications, and IT industries have now joined 5GAA. The Day-one function profile and system profile have been completed in 2018. Based on the proposed profiles, the interoperability among different systems can be enabled with the layered approach to facilitate the functional and performance verification.

In NGMN (Next Generation Mobile Networks), founded by the leading international mobile network operators in 2006, V2X task force was established in July 2016 to cooperate with the automotive industry. The views of operators on LTE-based V2X and DSRC were promoted and the trial results can be shared to reduce the time-to-market of LTE V2X [16].

Different countries and regions such as the United States, Europe, Japan and South Korea have been carrying out pilot areas construction, field tests, technical capabilities verifications of C-V2X in actual operation, and further accelerating the industrialization of intelligent connected vehicles (ICV). For example, the major activities include:

In Oct. 2017, Ford, Nokia, AT&T and Qualcomm started the first C-V2X regional trial in San Diego. The multi-OEM (Original Equipment Manufacturer) demos have been implemented and the interoperability has been verified.

In June 2018, working with Ford, Panasonic and Qualcomm, the Colorado Department of Transportation announced to be the first C-V2X technology deployment to show the critical step toward commercialization.

In 2018, the C-V2X trial in Japan was carried out with NTT Docomo, Nissan, Continental, Ericsson, and Qualcomm to validate and demonstrate the benefits of C-V2X. The key finding and observations, of C-V2X PC5 direct communication basic performance and Uu LTE network communications performance, have been summarized to show the benefits of C-V2X direct communications and network-based communications [17].

In April 2019, the first 5GAA C-V2X plugfest was held in Europe to demonstrate the high-level interoperability among the participating companies. The following 5GAA plugfest is undergoing with the collaboration of the participants [18].

### III. C-V2X IN CHINA

Having achieved fruitful results of C-V2X progress, China has carried out all-round layout and promotion in terms of policy planning, research and development of standards and technologies, industry landing, and field trial related to vehicle networking.

#### A. Policies

In order to further accelerate the innovation and development of ICV and strengthen the coordination, in Sep. 2017, the Special Committee of ICV Industry was established to be responsible for administrations including: organizing the top-level development of plans and policies, coordinating and solving the major problems in the development of ICV, supervising the implementation of relevant work, and promoting the industrial development. At the second meeting, in Nov. 2018, four national standardization associations had jointly signed the framework agreements on strengthening the cooperation in C-V2X standardization for automotive, intelligent transportation, communications and traffic management.

To accelerate the industrialization of C-V2X, the spectrum planning for LTE-V2X PC5 was officially issued by MIIT (Ministry of Industry and Information Technology) of China in Oct. 2018. The band of 5905 to 5925 MHz is dedicated to LTE-V2X [19]. The license of deployment and operation of LTE-V2X have been approved by MIIT to Hainan Province and Tianjin to conduct the on-the-ground field test.

In Dec. 2018, MIIT published the roadmap with "The industry promotion action plan for the intelligent and connected vehicles". The Action Plan will play the key role of policy leadership and achieve the goal of high-quality development of ICV industry in two stages. In the first stage, by 2020, the breakthroughs in the cross-industry integration of the vehicular networking industry will be achieved. LTE-V2X will be deployed in specific scenarios with a certain scale and the user penetration rate will be increased to be more than 30%. In the second stage, after 2020, the technological innovation, standards, infrastructures, application and security systems will be fully completed. The high level of ICVs can realize the automated driving and NR-V2X will gradually achieve the large-scale commercialization [20].

#### B. Standards

For the standardization of ICV, in order to give full play to the top-level design, and the leading and exemplary role, the guidelines for standard architecture of ICV (Part II: ICV) was released with four parts by MIIT in Dec. 2017 [21]. The ICV standards system structure is presented in Fig.2.

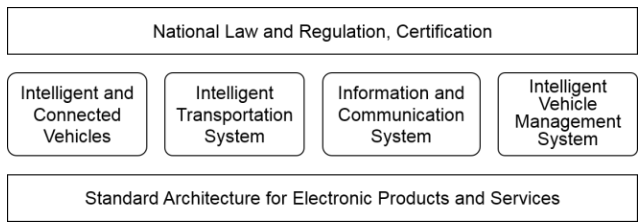

Fig. 2. The standard architecture of intelligent and connected vehicles (ICVs)

Merging the different requirements from automotive, transportation, communications industries, and traffic management, the standards organizations are conducting the standardization of layered stack to satisfy the realistic Chinese ICV requirements [22], [23]. As shown in Fig. 3, the core technical standards of LTE-V2X, including access layer, network layer, security layer, facility layer and application layer, have been completed in China. The application layer supports the diverse V2X applications with the defined message format in facility layer. Focusing on the functionality of the certification management, signature, verifying signature, encryption and decryption, the security layer is designed under the facility layer and above the network layer and access layer. The network layer and access layer support the V2X communications with Uu interface and PC5 interface with corresponding networking connections.

#### C. Interoperability Testing

In Nov. 2018, the world's first LTE-V2X triple-cross

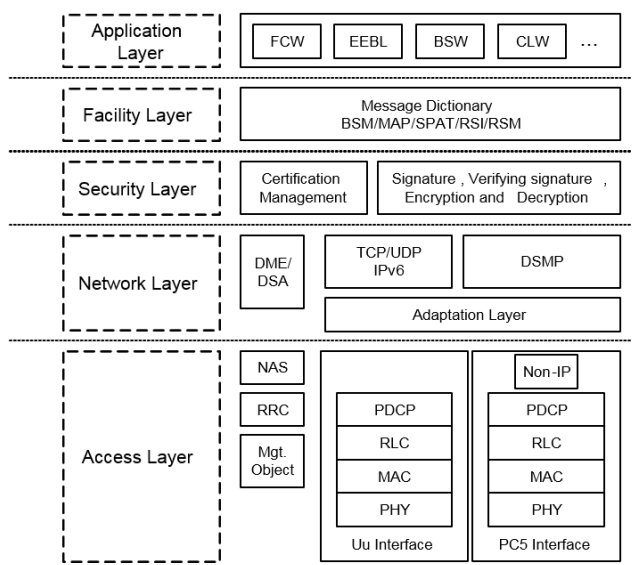

Fig. 3. The Chinese LTE-V2X standards system

FCW: Forward Collision Warning
EEBL: Emergency Electronic Brake Lights
BSW: Blind Spot Warning
CLW: Control Loss Warning
BSM: Basic Safety Message
MAP: Map Data
SPAT: Signal Phase and Time
RSI: Road Side Information
RSM: Road Safety Message
DME: DSRC Management Entity
DSA: DSRC Service Advertisement
TCP: Transmission Control Protocol
UDP: User Datagram Protocol
DSMP: DSRC Short Message Protocol
NAS: Non-Access Stratum
RRC: Radio Resource Control
PDCP: Packet Data Convergence Protocol
RLC: Radio Link Control
MAC: Medium Access Control
PHY: Physical Layer

interoperability trial was conducted by the CAICV (China Industry Innovation Alliance for the Intelligent and Connected Vehicles), IMT-2020 PG (International Mobile Telecommunications 2020 Promotion Group) C-V2X task force and SIAC (Shanghai International Automobile City). The cross-chipset/module, cross-TBOX (Telematics BOX), and cross-OEM LTE-V2X interoperability test is based on the Chinese V2X application layer standards and utilizing LTE-V2X Mode 4 as the communication mode. The interoperability with full protocol stack standards including the application layer, network layer, and access layer is verified to showcase the maturity and readiness of the ICV industry. With the basis of effective verification, LTE-V2X large-scale application deployment can be promoted and the collaborative industrial ecosystem can be constructed [24].

Based on the Chinese LTE-V2X security standards and the experience of the triple-cross interoperability trial, the security verification and demonstration of LTE-V2X communications will be conducted in Oct. 2019. The LTE-V2X security solutions will be showcased to promote the establishment of the secure and reliable LTE-V2X application environments.

### D. China ICV Development

In order to promote the C-V2X industry, MIIT and MT (Ministry of Transportation) cooperate with the multi local governments to promote the construction of Chinese pilot areas of ICV.

MIIT supported Beijing-Hebei, Chongqing, Zhejiang, Jilin, Hubei and other provinces to set up the C-V2X pilot areas to carry out the cross-industry demonstrations of the intelligent vehicles and transportations based on the broadband mobile Internet with applications, testing and verifications. The National Smart Transportation Comprehensive Test Base in Wuxi, Jiangsu Province was constructed and the pilot area is built with the deployment of LTE-V2X network in open road. The SIAC established the pilot area for National Intelligent Network Automobile (Shanghai) with LTE-V2X eNB and set up the C-V2X Server data center platform. The pilot areas are formed with the promotion of the verification of functionalities, performance and interoperability of LTE-V2X to build the foundation for the subsequent large-scale trial and the future industrialization and commercialization. Comparing to the USA., Europe and other countries and regions, with the advanced regulatory trends of C-V2X in China, China have led the global C-V2X development and entered the stage of rapid development to promote the deployment and promote the integration of 5G and ICV.

Besides the construction of pilot areas in China, the National major projects have been conducted to encourage the killer applications of ICV and promote the industrialization.

The special implementation program of "Tech-driven Winter Olympics" has been carried out for the 2022 Winter Olympic Games of Beijing. The cutting-edge technologies such as 5G and C-V2X will be massively adopted to support the 2022 Winter Olympic Games in Beijing.

### E. Industrial Practice

In China, the ecosystem of C-V2X is thriving and a complete industrial chain for C-V2X has been formed. For communication part, CATT/Datang launched the LTE-V2X commercial communication module DMD31, Huawei launched Balong 765, and Qualcomm launched 9150 chipset. For terminals, many companies, such as Neusoft, Nebula, Transinfo, and Quectel launched the OBU and RSU terminals based on the communication module or chipset. For applications, many OEMs and IT companies developed safety and efficient applications to support V2X services. In Apr. 2019, 13 Chinese OEMS jointly made an announcement that they will support volume production of C-V2X vehicles from the second half of 2020 to the first half of 2021.

CATT/Datang, as one of the leading research and development companies in C-V2X field, has conducted lots of industrial practices together with the automotive manufacturers.

#### 1) Public highway testing near Beijing

In May. 2018, Ford collaborated with CATT/Datang to conduct the C-V2X public road testing in the highway between Beijing and Tianjin. The vehicles are driven at 80 km/h, 100 km/h and 120 km/h, and the distance are maintained at 200 m, 400 m, and 600 m for about five minutes. The host vehicle and remote vehicle are separated with changing gap to verify the PER (Packet Error Rate) of LTE-V2X and IEEE 802.11p. The testing results are summarized in Fig. 4 which are very encouraging to show the benefit of LTE-V2X over IEEE 802.11p.

In Fig. 4, comparing with IEEE 802.11p, the communication range of LTE-V2X is about 600 m, which is about 200 m longer than IEEE 802.11p at the PER of 10% [25].

#### 2) Commercialization-oriented intelligent bus

As the first C-V2X based commercialization-oriented intelligent bus project in China, CATT/Datang developed the intelligent bus system with King Long which is the manufacturer of the BRT (Bus Rapid Transit) buses in Xiamen. The test was conducted along the commercial BRT route, which is the only commercial semi-open bus route in China.

The network architecture of the intelligent bus system is shown in Fig. 5. The C-V2X devices are deployed on the BRT

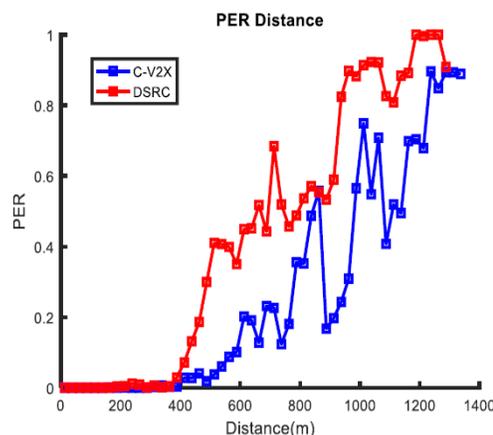

Fig. 4. Public highway testing of LTE-V2X conducted by Ford and CATT/Datang.

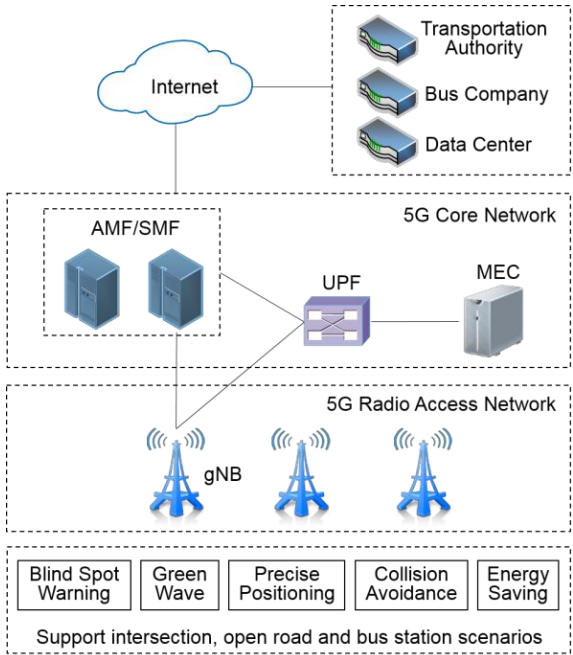

Fig. 5. The system architecture of the intelligent bus system in Xiamen

UPF: User Plane Function
AMF: Access and Mobility Management Function
SMF: Session Management Function
gNB: next generation Evolved Node B

buses and RSUs. 5G NR cellular network covers the whole BRT line. With MEC, V2V and V2I communications can provide the cooperative perception with low latency, high reliability and high throughput performance.

Supported by the intelligent bus system, some vehicle-road cooperative applications are verified, including: precise parking at bus stations, collision avoidance with NLOS (non Line-of-Sight) communication, information exchanging at intersection with NLOS communication, traffic signal adjusting and prioritized public bus, and optimal driving strategy.

The first stage verification of 3000 kilometers has been completed. The results show that the reliability of LTE-V2X and the stability of the bus with the LTE-V2X devices are guaranteed, and the fuel consumption can be reduced around 5%-10%.

## IV. THE TECHNICAL TRENDS AND CHALLENGES OF C-V2X APPLICATIONS

By providing information exchange and environment perception capability, C-V2X is the important enabling technology from single-vehicle intelligence to connected intelligence. If the vehicles rely on single vehicle sensing (often radars, lidars, cameras), intelligence (often on-board computing platform) and communication (often 4G communication modules), vehicles will suffer from the defects including: 1) Limited perception capability: this is mainly due to only LOS (Line-of-Sight) sensing range, non-robust perception in case of bad weather and sharp change of light, as well as difficulties in temporal and spatial synchronization. 2) Limited computation capability: all the complex computation tasks are executed on the on-board computation platform, which has very high price yet limited processing capability, thus preventing the volume deployment. 3) Limited communication capability: Using 4G communication modules can not satisfy the low latency and high reliability requirements of road safety applications, and cannot provide enough data rate for HDM (High Definition Map), VR (Virtual Reality) and AR (Augmented Reality) applications. Therefore, single-vehicle intelligence cannot realize completely autonomous driving.

C-V2X, together with other 5G network technologies such as MEC, can enable the evolution from single-vehicle intelligence to connected vehicle intelligence, i.e., vehicles and MEC hosts are connected to complete cooperative decision and control. C-V2X provides low latency and high reliability communications for data delivery, transmission of computation tasks, decision results and control instructions. Cooperative perception becomes possible by using V2V, V2I, V2P, V2N communications between the vehicles and other different elements along the roads. Benefited from the fundamental communication capabilities of C-V2X, MEC can provide powerful cooperative computing/storage capability, and enlarge information dissemination range.

Accordingly, it is envisioned that the applications of C-V2X

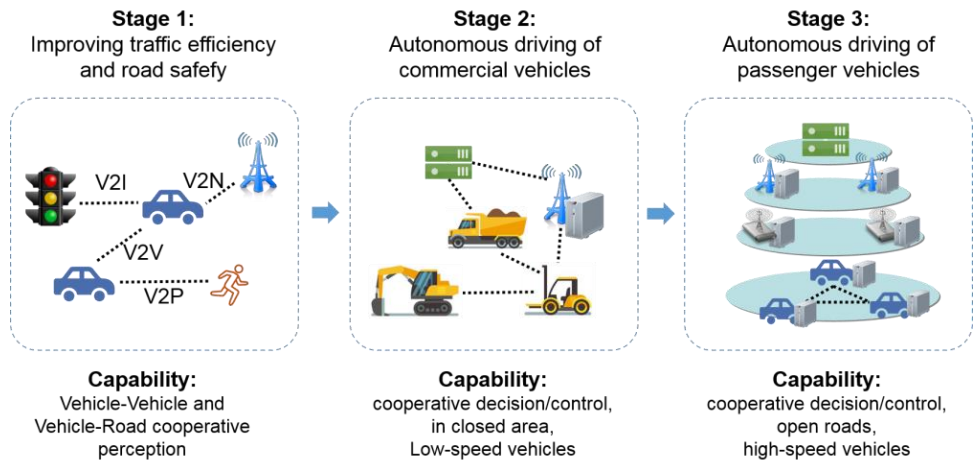

Fig. 6. Envisioned Application Phases of C-V2X

will experience the following three phases, as shown in Fig. 6.

Phase 1: Improving traffic efficiency and road safety. In this phase, C-V2X provides Vehicle-Vehicle and Vehicle-Road cooperative perception capabilities.

Phase 2: Autonomous driving of commercial vehicles (such as heavy-duty truck). Based on the MEC capability deployed in enclosed areas such as the industrial parks, harbors, wharfs, mining areas etc., C-V2X provides cooperative decision and control capabilities for the commercial vehicles driving with low speed in those enclosed areas.

Phase 3: Autonomous driving of passenger vehicles. When C-V2X and MEC are deployed widely, cooperative decision and control capabilities are provided for high-speed passenger vehicles.

In the future development of C-V2X and its applications in autonomous driving and intelligent transportation systems, the following technical challenges should be addressed.

### A. Integration with MEC

MEC is one of the promising technologies of 5G. MEC pushes the computation and storage resources to the edge of the network, which is "closer" to the users, and thus can augment the capability of mobile terminals. At the same time, MEC can meet the stringent requirements of low latency, high reliability, transmission efficiency, and deployment flexibility.

Integration of C-V2X and MEC can provide communication-computing-storage convergence for vehicular networks, thus provide support for the prospective use cases such as traffic signal control, congestion identification and analysis, real-time HDM loading, path planning, heterogeneous data fusion etc. [4, 26, 27]. In these advanced applications, computing-intensive or data-intensive tasks such as big data analysis, data mining and deep learning are required. The integration of C-V2X and MEC is an optimal choice in these use cases to avoid high communication and computation delay of cloud-centric solutions, and thus can meet the low latency and high reliability requirements of V2X applications.

In order to achieve efficient convergence of the communication-computing-storage resources, the technical challenges in the integration of C-V2X and MEC include: dynamical deployment and joint scheduling of various resources, computation offloading decision, communication handover and computation migration in high mobility scenarios. In addition, the interoperability and compliance of the C-V2X technology is essential, several key standardization organizations (such as ETSI, 3GPP, and 5GAA) are focusing on the MEC technologies, testing and verifications to promote the communications among the participating V2X applications in different MEC systems.

### B. Channel Modeling

In vehicular communications, channel modeling is the foundation for communication system design and performance evaluation. However, because of the unique propagation conditions, V2V channels exhibit significantly different propagation characteristics compared to the cellular communication channels [28]. In V2V communications, both the transmitter and the receiver are with high mobility. In addition, there may exist a large number of surrounding scatterers which could also be moving. Accordingly, the radio channel of vehicular networks generally has the following special characteristics [28, 29, 30, 31]: large and time-varying Doppler shifts, deep fading (i.e., presence of "worse than Rayleigh" fading in a significant percentage of cases), as well as the non-stationarity feature in space-time-frequency domains.

In summary, the challenges on channel modeling in vehicular networks include: hybrid statistical geometric modeling, the time-varying vehicular channels measurements and modeling at mmWave (millimeter wave) band, non-stationary channel characterization and modeling in space-time-frequency domains, tracking and dynamic clustering of multipath components, as well as channel prediction based on machine learning.

### C. 5G enhancement based high-accuracy positioning

Vehicle driving decision-making requires very high accuracy positioning. For example, a 100 km/h vehicle travels 28 meters per second. Once the positioning deviation is large, the safety of the vehicle is difficult to guarantee. Only by relying on high-accuracy positioning, the safety of driving and decision-making can be ensured. Because 5G adopts multi-antenna technology, positioning capability has been greatly improved. In 5G network, high definition maps can be updated quickly. With RTK (Real-time kinematic) positioning and inertial navigation positioning technology, the accuracy of vehicle positioning can be on centimeter level. At the same time, using C-V2X, the location information between vehicles can be exchanged in real time to ensure the relative positioning ability of vehicles in multi-vehicle environment. In a word, under the condition of 5G C-V2X, the integration of multiple positioning technologies can be used to achieve reliable, stable and high- accuracy positioning.

Although 5G-based high-precision positioning has good prospects, but in the current standard of 3GPP, there is no clear definition of 5G network high-precision positioning standards, and RTK is not suitable for large-scale applications because of its high price. In addition, there is a lack of feasible relative positioning technology. Relative positioning technology can greatly reduce the cost of vehicle positioning, the RSU or roadside 5G eNB can be used to form high-precision positioning of vehicles in the region.

### D. The integration of radar and C-V2X communications

Traditionally, radar and communications have typically been developed in isolation [32]. In vehicular networks, both data exchange and target detection are required. And because both most automotive radars and future C-V2X communication systems work at millimeter wave band, the radar-communication integration and co-design becomes one of the promising and challenging technologies[32, 33]. By combining spectral and hardware resources, such an integration system has the advantages of low cost, compact size, and improved spectrum efficiency [33]. The detection results of different radars can also be shared and jointly processed based on the communication capability, therefore to improve detection accuracy and to achieve comprehensive environment sensing for automotive applications.

The radar-communication integration and co-design still faces various technical challenges. C-V2X communication and radar have different requirements on modulation, bandwidth and circuits. Therefore, joint waveform design and signal separation at mmWave band, radar echo orientation based on communication waveform, extremely high accuracy time synchronization are needed to achieve the integration of C-V2X communications and detection. On the other hand, detection result sharing and joint processing among multiple automotive radars still need to tackle the problems such as high data rate and low latency transmission, coordinate alignment among communication peers, and communication/radar beamforming etc.

## V. CONCLUSION

C-V2X provides low latency, high reliability and high throughput V2X communications by leveraging and enhancing the current cellular systems. It evolves from LTE-V2X to NR-V2X. LTE-V2X provides information exchange capability for basic road safety applications, while NR-V2X will coexist with LTE-V2X, aiming to support advanced applications in autonomous driving and intelligent transportation systems.

This paper presents a vision of C-V2X. According to the evolution path from LTE-V2X to NR-V2X, the requirements of both road safety and advanced V2X applications, the centralized/distributed architecture, the key technologies of C-V2X are introduced. Standards research in different international standardization organizations, global deployment, field testing and technical verifications are also summarized. Especially, based on the continually and actively promotion of C-V2X research, field testing and development in China, the related works, and the progresses in interoperability testing and industrial practice are also presented.

As the technical trends of C-V2X, together with other 5G technologies such as MEC, C-V2X can enable the evolution from single-vehicle intelligence to connected vehicle intelligence, which is critical for autonomous driving and intelligent transportation systems. Accordingly, in this paper, it is envisioned that the application of C-V2X will experience three phases: improving traffic efficiency and road safety, autonomous driving for commercial vehicles, and autonomous driving for passenger vehicles. In addition, the technical challenges of integration with MEC, channel modeling, 5G enhancement based high-accuracy positioning, and the integration of radar and C-V2X communications are analyzed.

ACKNOWLEDGMENT

The authors would like to give special thanks to Prof. Dake Liu of Beijing Institute of Technology for his kind review and revisions, thanks to Mr. Jiayi Fang from CATT, Prof. Xiang Cheng from Peking University, Prof. Jianhua Zhang from Beijing University of Posts and Telecommunications, and Prof. Ruise He from Beijing Jiaotong University for their kind advices.

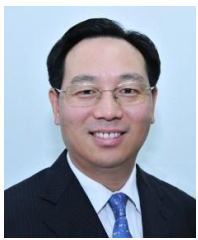 **Shanzhi Chen**(SM'04, F'20) received the bachelor's degree from Xidian University in 1991 and the Ph.D. degree from the Beijing University of Posts and Telecommunications, China, in 1997. He joined the Datang Telecom Technology and Industry Group and the China Academy of Telecommunication Technology (CATT) in 1994, and has been serving as the EVP of Research and Development since 2008. He is currently the Director of the State Key Laboratory of Wireless Mobile Communications, CATT, where he conducted research and standardization on 4G TD-LTE and 5G. He has authored and co-authored four books, 17 book chapters, more than 100 journal papers, 50 conference papers, and over 50 patents in these areas. He has contributed to the design, standardization, and development of 4G TD-LTE and 5G mobile communication systems. His current research interests include 5G mobile communications, network architectures, vehicular communication networks, and Internet of Things. He served as a member and a TPC Chair of many international conferences. His achievements have received multiple top awards and honors by China central government, especially the Grand Prize of the National Award for Scientific and Technological Progress, China, in 2016 (the highest Prize in China). He is the Area Editor of the IEEE INTERNETOFTHINGS, the Editor of the IEEE NETWORK.

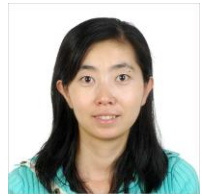 **Jinling HU** (hujinling@catt.cn) received her master's degree from Beihang University in 1999. She is the Chief Expert at Gohigh Data networks technology Co. LTD of CATT, where she works on research of key technologies in next generation mobile communications.

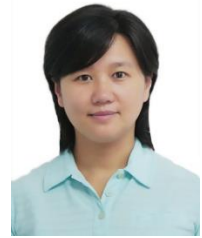 **Yan SHI** (shiyan@bupt.edu.cn) received her Ph.D. degree from Beijing University of Posts and Telecommunications (BUPT) in 2007. She is currently an associated professor of the State Key Laboratory of Networking and Switching Technology, BUPT. Her current research interests include network architecture evolution, protocol design and performance optimization of future networks and mobile computing, especially mobility management technology.

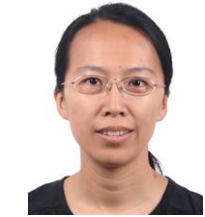 **Li ZHAO**(zhaoli@catt.cn) received her master's degree from Beijing University of Posts and Telecommunications (BUPT) in 2004. She is currently a senior engineer at Gohigh Data networks technology Co. LTD of CATT. Her current research interests focus on vehicular networking and cellular high layer protocol.

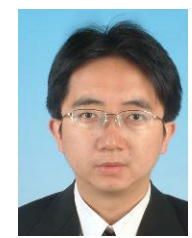 **Wen LI**(liwen@datangmobile.cn) received his master's degree from the Graduate School of Chinese Academy of Sciences. He is vice president of 5G product line with the Datang Mobile in business application research direction.